\documentclass[reprint,amsmath,amssymb,aps,prl,floatfix]{revtex4-2}

\usepackage[title]{appendix}
\usepackage{xcolor}
\usepackage{verbatim}
\usepackage{graphicx}
\usepackage{bm}
\usepackage{lineno}
\usepackage{setspace}
\usepackage{todonotes}
\usepackage{cuted}
\usepackage{soul}
\usepackage{todonotes}
\usepackage{tabularx}
\usepackage{array}
\usepackage{booktabs}

\newcolumntype{C}{>{\centering\arraybackslash}X}

\begin{document}

\preprint{}

\title{Quasi-normal-mode signatures of periodicity and hyperuniformity}

\author{V. Romero-Garc\'ia$^1$}\email{virogar1@mat.upv.es}
\author{M. Martí-Sabaté$^1$}
\author{V. F. Dal Poggetto$^1$}
\author{L.M. García-Raffi$^1$}
\author{M. L\'azaro$^1$}

\affiliation{$^1$Instituto Universitario de Matem\'atica Pura y Aplicada, Universitat Polit\`ecnica de Val\`encia, Camino de Vera, s/n 46022 Val\`encia, Spain}

\date{\today}

\begin{abstract}
Hyperuniform materials occupy an intermediate regime between periodic order and conventional disorder, characterized by a suppression of long-wavelength density fluctuations, commonly quantified by the structure factor. This characterization, however, relies on the statistical properties of an extended configuration or supercell representation and does not directly describe the resonant response of a finite structure, as typically encountered in scattering experiments. Here, we introduce the quasi-normal-mode spectrum as a complementary framework for characterizing the wave properties of finite stealthy hyperuniform materials. We analyze the complex eigenfrequencies of both periodic and stealthy hyperuniform structures over a range of the stealthiness parameter ($\chi$), and investigate how the spatial organization of scatterers is encoded in both frequency and spatial profiles of their quasi-normal modes. We find systematic changes in the quasi-normal modes spectrum as the degree of hyperuniformity is varied, revealing signatures that are not captured solely by the structure factor. In particular, the evolution of the resonant states provides a direct connection between structural correlations and the scattering response of finite systems. Our results establish quasi-normal modes as a real-space and open-system characterization of hyperuniformity and provide a unified framework for comparing discrete heterogeneous materials from the perspective of their scattering and resonant dynamics.
\end{abstract}

\maketitle


\section{Introduction}
Quasi-normal modes (QNMs) provide a natural framework for describing the resonant properties of open wave systems~\cite{Ching98, Evans94, Hsu13, Hsu16}. In contrast to the real-frequency eigenmodes of closed systems characterized by self-adjoint operators, QNMs satisfy outgoing-wave boundary conditions and consequently possess complex eigenfrequencies, whose real and imaginary parts characterize, respectively, the resonance frequency and its radiative decay rate~\cite{Pagneux13, Koshelev18, Koshelev23}. The QNM framework has been extensively developed across a remarkably broad range of physical systems, including optical~\cite{Lalane18, Leung94, Sauvan21}, acoustic~\cite{Sabate23, Sabate24, Huang21}, water~\cite{Linton07}, and elastic waves~\cite{Laude23, Wen25, Laude26, He25, vial2024quasinormal,sabate2026fabry}, as well as gravitational physics and cosmology~\cite{VISHVESHWARA70, Chandrasekhar75, Nollert92, Kokkotas99, Motohashi25}. In the latter context, QNMs provide a fundamental description of the damped oscillations of black holes and compact astrophysical objects, while in wave-scattering systems they are naturally interpreted as poles of the scattering response of open structures~\cite{Ching98}. In both cases, the complex eigenfrequency simultaneously encodes the characteristic oscillation frequency and the rate at which the mode decays through its available radiation channels. Beyond providing the resonance spectrum, the associated eigenvectors give access to the spatial structure and localization of the resonant states. This distinction is particularly important for finite disordered systems, for which the conventional band structure of periodic media is no longer sufficient to characterize the spectrum of an individual scattering realization. Recent developments have further established QNMs as a powerful basis for describing resonant scattering and open-system dynamics~\cite{Rotter15, Lyapina15}, including systems with complex geometries~\cite{Christopoulos24}, intrinsic losses~\cite{Krasikova24}, and non-Hermitian spectral properties~\cite{Hatano21, Hashemi22}.

Hyperuniform materials provide a remarkable class of disordered structures in which the absence of conventional translational order coexists with strong suppression of long-wavelength density fluctuations~\cite{Torquato13, Torquato16, Torquato18}. This hidden structural order is commonly characterized by the structure factor~\cite{ashcroft76} and becomes particularly restrictive in stealthy hyperuniform systems~\cite{Torquato15,Torquato15_1, Torquato15_2}. Such correlated disorder has been shown to produce wave phenomena that are qualitatively different from those of uncorrelated random media, including isotropic and complete band gaps~\cite{Florescu09,Froufe16,Gkantzounis17,Aubry20,Cheron22_1, Cheron22_2,Rohfritsch21}, unusual transport regimes~\cite{Romero19,Romero21, Kuznetsova21,Kuznetsova24}, and modified localization properties~\cite{Man13,   Alhaitz23, Karcher24, Simon24, Barsukova26, Valier-Brasier26}. In particular, hyperuniformity provides a route to control wave propagation without relying on conventional periodic order, making these systems an important platform for studying the interplay between structural correlations and wave physics. However, the spectral characterization of hyperuniform structures has predominantly focused on bulk quantities, such as the structure factor or density of states, among others. These descriptors are naturally associated with extended or infinite systems and do not directly resolve the complex resonant spectrum of a finite open realization. Consequently, the effects of hyperuniform spatial correlations on the quasi-normal modes of individual finite structures, including their complex frequencies and spatial localization remain comparatively unexplored.

We analyze these properties using a thin elastic plate decorated with point masses as a platform of investigation, considered as a minimal model of a heterogeneous wave-bearing medium. It provides a direct, physically transparent way to introduce spatial disorder while keeping the properties of the host medium homogeneous. The point masses modify the inertial response of the plate locally and therefore act as scatterers for flexural waves, with their positions defining the spatial correlations of the resulting medium. By prescribing these positions according to periodic or stealthy hyperuniform point patterns, the same mechanical system can be used to systematically investigate the effect of structural order and long-range correlations on the resonant properties of the medium. This approach is particularly convenient for the present study because it separates the effects of the underlying spatial organization from those associated with variations in the host plate material properties. Moreover, the point-mass representation leads to a well-defined biharmonic wave equation with localized inertial perturbations, providing a simple framework in which the quasi-normal modes of an open, finite structure can be directly identified and related to its scattering response.

In this work, we investigate the resonant properties of finite periodic and hyperuniform media from the perspective of their quasi-normal modes. We consider a thin elastic plate decorated with point masses, and formulate the problem using a multiple-scattering approach. This allows us to determine the complex-frequency spectrum of the open system and to establish a direct connection between the spatial organization of the scatterers and the resonant response of the structure. We systematically analyze configurations ranging from periodic order to hyperuniform states characterized by different values of the stealthiness parameter $\chi$~\cite{Torquato13}. Rather than relying exclusively on the structure factor, which only characterizes the long-wavelength density fluctuations of an extended point pattern, we use the distribution of QNMs to characterize the dynamical response of the finite scattering system. By examining the evolution of the QNM frequencies, linewidths, and spatial profiles across the different classes of structures, we identify spectral signatures associated with the suppression of long-range density fluctuations and demonstrate the potential of QNMs as a complementary dynamical descriptor of hyperuniformity.

\section{Theoretical framework}
\subsection{Self-adjoint and non-self-adjoint operators}

Consider an operator $\mathcal{L}$ defined on a Hilbert space with inner product $\langle\cdot,\cdot\rangle$, whose adjoint $\mathcal{L}^{\dagger}$ is defined through the relation $\langle u,\mathcal{L}v\rangle = \langle \mathcal{L}^{\dagger}u,v\rangle$ for all functions $u$ and $v$ belonging to the appropriate domains~\cite{schmuedgen12}. Importantly, the adjoint is determined not only by the differential expression defining $\mathcal{L}$, but also by its domain and boundary conditions, typically obtained as the result of integration by parts---or, in higher-dimensional domains, the corresponding Green identities---. Such operator is said to be \emph{self-adjoint} when $\mathcal{L}=\mathcal{L}^{\dagger}$, including equality of their domains. For example, the Laplacian $-\nabla^2$ and the biharmonic $\nabla^4$ operators can both define self-adjoint eigenvalue problems when supplemented with suitable conservative boundary conditions~\cite{simon80}. In this case, the eigenvalues are real and the corresponding eigenfunctions can be chosen to form an orthogonal set under the appropriate inner product. The resulting modes are the conventional \emph{normal modes} of a closed system~\cite{landauLifshitz76, courantHilbert89}.

This situation changes fundamentally for an open (unbounded) scattering problem. In this case, the system exchanges energy with the surrounding medium through radiation, and the physical boundary condition is an outgoing-wave rather than a conservative boundary condition. Consequently, the boundary terms generated by integration by parts do not, in general, vanish, and the operator equipped with the outgoing-wave boundary condition is non-self-adjoint. The corresponding eigenvalues are therefore generally complex, with their imaginary parts describing the finite lifetime associated with radiative decay~\cite{Ching98, sauvan22}. The resulting eigenfunctions are not conventional normal modes but QNMs. Thus, although the underlying differential operator remains the same Laplacian or biharmonic operator, changing the boundary conditions alters its domain and transforms the spectral problem from a self-adjoint eigenvalue problem into a non-self-adjoint one. In scattering problems, the appropriate terminology is consequently that of QNMs, whose complex eigenfrequencies characterize the resonant response and encode both the resonance frequency and its radiative linewidth.


\subsection{Multiple Scattering for flexural waves in plates decorated with point masses}

Considering a thin, homogeneous, and isotropic solid plate under the Kirchhoff-Love plate theory, and a harmonic temporal dependence of type $e^{-i\omega t} $, with $i^2=-1$, $\omega$ the angular frequency, and $t$ the time, the flexural displacement $W(\mathbf{x})$ is described by~\cite{graff91}
\begin{equation}
\left( \nabla^4 - k^4 \right) W(\mathbf{r}) = 0,
\end{equation}
where $k^4 = \frac{\rho h \omega^2}{D}$, with \(D = \dfrac{E h^3}{12(1-\nu^2)}\), \(\rho\), \(h\), $E$ and $\nu$ are the flexural wave number, the flexural rigidity, the density, thickness, the Young's modulus, and the Poisson ratio of the plate material, respectively.

Consider a point-mass (scatterer) placed at position ${\bf R}_{\alpha}$. The equation of motion of the vertical displacement in the plate, $W_1$, reads as~\cite{torrent21, Lazaro26}
\begin{eqnarray} (\nabla^4-k^4)W_1(\textbf{r})=t_\alpha W_1(\textbf{R}_\alpha)\delta(\textbf{r}-\textbf{R}_\alpha), \label{eq:inhomogeneousEqMotion}\end{eqnarray}
where 
\begin{eqnarray}
t_{\alpha}=
\frac{\omega^2 m_{\alpha}}{D}
\label{eq:ta_mas}
\end{eqnarray}
is the strength of each point-like scatterer
and it is the only quantity that contains information about
its physical properties; $m_{\alpha}$ represents the mass of the scatterer, ans $\delta$ denotes the Kronecker delta

Based on the Green's function method, we can solve the Eq. \eqref{eq:inhomogeneousEqMotion} as follows
\begin{eqnarray}
W_1({\bf r})
=
\psi_0({\bf r})
+ T_{\alpha} \, \psi_0({\bf R}_{\alpha}) \,
G({\bf r}-{\bf R}_{\alpha}),
\end{eqnarray}
where the Green's function is
\begin{eqnarray}
G({\bf r})=
\frac{i}{8k^2}
\left[
H_0(kr)+\frac{2i}{\pi}K_0(kr)
\right],
\end{eqnarray}
where $H_0(kr)$ is the  order zero Hankel function of the first kind and $K_0(kr)$ is the  order zero modified Bessel function of the second kind, the incident wave is represented by $\psi_0$, and $T_{\alpha}$ is related to the strength of the scatterers through,
\begin{eqnarray}
T_{\alpha}
&=&
\frac{t_{\alpha}}{1-i\frac{t_{\alpha}}{8k^2}}.
\end{eqnarray}

Consider now $N$ point masses placed at positions ${\bf R}_{\alpha}$, with $\alpha=1,\ldots,N$. The wave dynamics of this mass-loaded plate can be formulated within a multiple-scattering framework, in which each point mass is regarded as an individual localized scatterer embedded in an otherwise homogeneous plate. The displacement field is expressed as the superposition of the incident field and the fields scattered by all point masses.
\begin{eqnarray}
    W(\mathbf{r}) =
\psi_{0}(\mathbf{r}) +
\sum_{j=1}^{N} \psi_{\mathrm{s}}^j(\mathbf{r}).
\end{eqnarray}

Importantly, the field acting on a given scatterer is not only the externally incident field but also the field generated by every other scatterer, 
\begin{eqnarray}
\psi_{\mathrm{e}}^j(\mathbf{r}) &=&
\psi_{0}(\mathbf{r})
+
\sum_{\substack{m=1 \\ m \neq j}}^{N}
\psi_{\mathrm{s}}^m(\mathbf{r}),
\end{eqnarray}
thereby accounting for all orders of multiple scattering and the collective interaction between the scatterers. These interactions are mediated by the Green's function of the homogeneous plate, which describes the propagation of flexural waves between two arbitrary positions. By using the multiple-scattering decomposition as in Refs.~\cite{torrent21, Lazaro26}, the resulting self-consistent system of equations is given by
\begin{eqnarray}
    \sum_{\beta=1}^N\left[\delta_{\alpha,\beta}t_{\beta}^{-1}-G({\bf R}_{\alpha}-{\bf R}_{\beta})\right]T_{\beta}\psi_{\mathrm{e}}({\bf R}_{\beta})=\psi_0({\bf R}_{\alpha}).
\end{eqnarray}
The solution of the previous system determines the amplitudes of the scattered fields at all point masses and, consequently, the full wave field of the structure. This formulation is particularly well suited to disordered and hyperuniform configurations because the individual scatterers and their positions enter explicitly into the governing equations, allowing the effect of spatial correlations to be directly connected to the collective resonant response.

At this stage it is important to  introduce the Foldy-Lax matrix~\cite{foldy1945multiple,lax1951multiple,lax1952multiple} as the matrix that relates the scattered amplitudes, $\psi_{\mathrm{e}}({\bf R}_{\beta})$, to the incident ones $\psi_0({\bf R}_{\alpha})$, and we denote it by $\bm{\mathcal{M}}$. Its coefficients are given by
\begin{eqnarray}
    \mathcal{M}_{\alpha\beta}=[\delta_{\alpha,\beta}t_{\beta}^{-1}-G({\bf R}_{\alpha}-{\bf R}_{\beta})]T_{\beta}.
    \label{Foldy}
\end{eqnarray}
The elements of the matrix depend only on the properties of the scatterers, on their spatial distribution, and on the frequency.

\subsection{Quasi-normal modes}

The resonant modes or QNMs of a passive scattering system are defined as the non-trivial solutions of the homogeneous problem with Sommerfeld radiation conditions, corresponding to complex frequencies $\omega_n$ such that
\begin{eqnarray}
\det\!\big[\bm{\mathcal{M}}(\omega_n)\big] = 0.
\end{eqnarray}
Each QNM is associated with a complex frequency ${\omega_n = \Omega_n - i\gamma_n}$, characterized by a resonance frequency $\Omega_n$ and a decay rate $\gamma_n$, with $\gamma_n>0$ for the time dependence $e^{-i\omega t}$. In practice, QNMs can be obtained numerically by searching for the zeros of $\det[\bm{\mathcal{M}}(\omega)]$ in the complex frequency plane. The associated modes correspond to the natural modes of region occupied by the finite number of scatterers. These modes, which mathematically correspond to zeros of $\det(\bm{\mathcal{M}}(\omega))$ in the complex frequency plane, manifest physically as resonances.

We note that the QNM spectrum exhibits the symmetry $\omega_n\leftrightarrow-\omega_n^*$ with respect to the imaginary-frequency axis~\cite{Ching98}, where $(\cdot)^*$ denotes the complex conjugate. This symmetry results from the combination of time-reversal invariance and the reality of the underlying wave equation. More specifically, because the Green's function is constructed from the outgoing fundamental solution, time reversal transforms an outgoing solution at $\omega$ into an incoming solution at $-\omega^*$. The symmetry of the QNM spectrum therefore involves reflection with respect to the imaginary axis rather than simple complex conjugation, the latter transforming an outgoing solution into an incoming one. The QNMs consequently occur in pairs related by $\omega_n\leftrightarrow-\omega_n^*$.

\subsection{Stealthy hyperuniform materials}

Hyperuniform materials are a class of statistically disordered structures characterized by the suppression of density fluctuations at large length scales~\cite{Torquato13, Torquato16, Torquato18}. Unlike conventional disordered systems, whose long-wavelength density fluctuations are typically comparable to those of an uncorrelated random distribution, a hyperuniform point pattern exhibits a structure factor~\cite{ashcroft76} that vanishes in the limit of zero wavenumber,
\begin{equation}
    S(\mathbf{q}) \rightarrow 0,
    \qquad |\mathbf{q}|\rightarrow 0,
\end{equation}
where the structure factor for a set of $N$ points located at positions $\mathbf{r}_j$, with $j=1,\ldots,N$, in an square domain of side $L$, is
\begin{equation}
S(\mathbf{q})
=
\frac{1}{N}
\left|
\sum_{j=1}^{N}
e^{-i\mathbf{q}\cdot\mathbf{r}_j}
\right|^2.
\end{equation}
with $\mathbf{q}$ the vector in the reciprocal space. For a periodically repeated square domain, the allowed reciprocal vectors are given by $\mathbf{q}= \frac{2\pi}{L}\mathbf{n}$, where $\mathbf{n}= (n_x,n_y)$ with $n_x,n_y\in\mathbb{Z}$. This metric indicates that density fluctuations are strongly suppressed over length scales much larger than the characteristic inter-particle distance, despite the absence of conventional translational periodicity. Hyperuniform systems constitute a distinct class of structures characterized by suppressed long-wavelength density fluctuations. They can exhibit strong spatial correlations and long-range order without possessing the translational periodicity of crystals, thereby providing a framework that encompasses both ordered and disordered forms of matter~\cite{Batten08, Morse24}. In wave systems, the suppression of long-wavelength density fluctuations characteristic of hyperuniformity can substantially alter the scattering of long-wavelength excitations, thereby modifying the spectral and transport properties of the medium~\cite{Florescu09,Leseur16, KimTorquato24}, commonly under the assumption of weak scattering~\cite{Lazaro26}.

A particularly important subclass is formed by stealthy hyperuniform structures ~\cite{Torquato15,Torquato15_1, Torquato15_2}. In these systems, the suppression of density fluctuations extends over a finite region of the reciprocal space, such that $S(\mathbf{q})=0$, $0<|\mathbf{q}|<K$, where $K$ denotes the stealthy cutoff. Consequently, density fluctuations associated with a finite range of long-wavelength spatial Fourier components are completely suppressed. The parameter $K$, or equivalently the dimensionless stealthiness parameter $\chi=\frac{M(K)}{d(N-1)}$, where $M(K)$ is the number of independent constrained ${\bf q}$-vectors inside the stealthy region and $d$ the dimension of the problem, is commonly used to characterize these configurations and controls the degree of structural correlations and the proximity of the system to crystalline order~\cite{Torquato15}. Increasing $\chi$ generally imposes stronger constraints on the point configuration and reduces the number of independent degrees of freedom. Stealthy hyperuniform structures therefore provide a convenient framework for continuously tuning the spatial correlations of a disordered material while preserving the absence of conventional periodic order.

In the current work, these structures are used to investigate how long-range spatial correlations, quantified by the stealthiness parameter $\chi$, are manifested in the complex-frequency spectrum of finite open systems and in the spatial character of their QNMs.

\subsection{Plane-wave expansion with the supercell approximation}

To characterize the spectral properties of the hyperuniform configurations, we also employ the plane-wave expansion (PWE) method by considering each finite realization as a supercell that is periodically repeated in space~\cite{Mirand23, Xiao12, Romero10_1, Romero10_2, Romero10, Romero11}. Although the original point distribution is aperiodic, the supercell approximation allows the configuration to be described within a Bloch framework by imposing periodicity over a length scale corresponding to the size of the finite realization. The material parameters are therefore expanded in reciprocal-space Fourier components of the supercell, and the governing wave equation is transformed into an eigenvalue problem for the Bloch wavevector. By scanning the first Brillouin zone of the periodically repeated supercell, the resulting eigenfrequencies provide an effective band structure associated with the selected hyperuniform realization.

\section{Results}

\subsection{Periodic systems and defect states}

\begin{figure*}
    \centering
    \includegraphics[width=\linewidth]{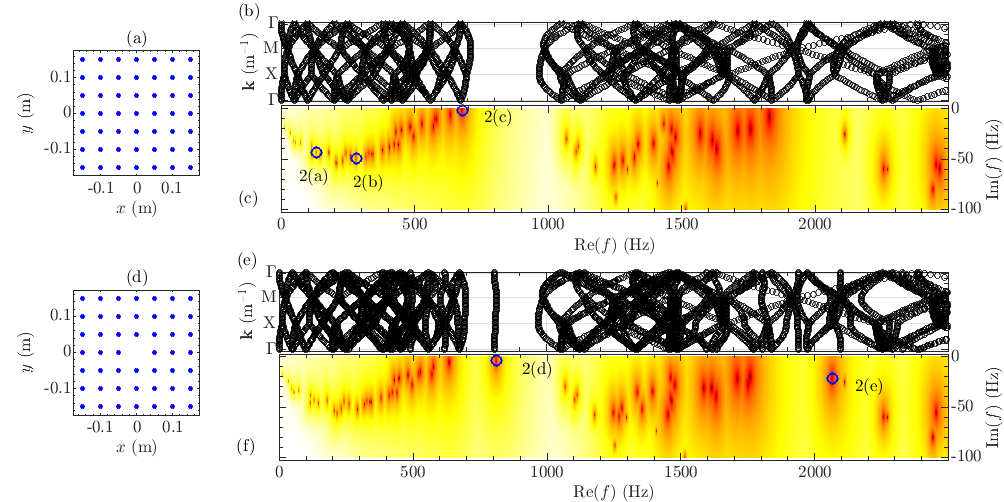}
    \caption{Comparison between the band structure of the periodic mass-loaded plate and the QNM spectrum of a finite $7\times7$ supercell. The point masses have mass $m=0.01\;\mathrm{kg}$ and are arranged on a square lattice with periodicity $a=0.05\;\mathrm{m}$. (a) Supercell used in the PWE calculation. (b) shows the dispersion relation of the infinite periodic structure, obtained from the self-adjoint eigenvalue problem by using the PWE with supercell approximation. (c) shows the QNM frequencies of the corresponding finite open supercell in the complex-frequency plane. The real-frequency pass bands are populated by QNMs, whereas the band-gap regions exhibit no QNM frequencies. The QNM spectrum displays the symmetry $\omega\leftrightarrow-\omega^*$ associated with time-reversal symmetry and the outgoing-wave Green's function. (d) Supercell for the PWE calculations of the crystal with a point defect at the center of the supercell. (e) Band strucutre of the defected crystal. The defect introduces a localized state inside the band gap as shown in the dispersion relation displayed in the upper panel. (f) The corresponding complex frequency plane; the defect mode appears as a QNM with a small imaginary part and consequently a high quality factor. Blue open circles show the zeros at which the modal displacement is described in Fig.~\ref{fig:fig2}.
    }
    \label{fig:fig1}
\end{figure*}

To establish a connection between the conventional band structure of periodic structures and the QNM spectrum, we first consider a $7\times7$ supercell of the periodic mass-loaded plate, with point masses of 0.01 kg arranged on a square lattice with lattice constant $a=0.05$ m, as shown in Fig.~\ref{fig:fig1}(a). 

The dispersion relation of the corresponding infinite periodic system is obtained from a self-adjoint eigenvalue problem using PWE for the case $\omega(k)$ and therefore consists of real eigenfrequencies, showing the pass bands and band gaps of the structure [see Fig.~\ref{fig:fig1}(b)]. We then consider the same $7\times7$ configuration as a finite open system and determine its QNMs. In contrast to the periodic problem, this one is not self-adjoint and therefore the eigenfrequencies of the open system are complex. Figure~\ref{fig:fig1}(c) shows the minimum eigenvalue of the Foldy-Lax matrix in terms of the complex frequency, Eq.~(\ref{Foldy}), in logarithmic scale to highlight the presence of the zeros, corresponding to the zeros of the determinant and as a consequence to the QNMs of the open system. 

The comparison between the two spectral descriptions reveals how the band structure of the infinite crystal is encoded in the resonant spectrum of the finite open system. As shown in Fig.~\ref{fig:fig1}(c), the real parts of the QNM frequencies are concentrated in the frequency intervals corresponding to the pass bands of the periodic system, where propagating states are supported. 
The QNM spectrum has a frequency distribution that retains the spectral fingerprints of the bulk band structure without requiring periodic boundary conditions. 

This correspondence becomes particularly apparent when a point defect is introduced at the center of the supercell. As shown in Fig.~\ref{fig:fig1}(d), the defect, created by removing the cental mass of the supercell, produces a localized state within the band gap. In the open-system formulation, this state appears as a QNM with a real part lying inside the bulk band gap and a very small negative imaginary part as shown in Fig.~\ref{fig:fig1}(f). The latter quantifies its weak coupling to the radiative continuum and corresponds to a long-lived resonance with a high quality factor, $Q=|\Omega_n/2\gamma_n|$. The defect-induced QNM therefore provides a direct dynamical signature of the localized state that is absent from the band structure of the crystal. We analyze this from the analysis of the calculated quality factors as shown in Tab.~\ref{tab:QNM_Q}.

We start by the modes at $\omega_n=2\pi(130.6-i43.5)$~rad/s and $\omega_n=2\pi(279.9-i49.3)$~rad/s. They have relatively large imaginary parts, corresponding to $Q\simeq1.5$ and $Q\simeq2.8$, respectively. These broad resonances are characteristic of strongly radiative states, for which the coupling to the surrounding continuum is sufficiently strong to produce short modal lifetimes. Their relatively small quality factors are consistent with their location within propagating spectral regions, where efficient coupling to extended states and, consequently, to the external radiation channels is possible.

In contrast, the QNM at $\omega_n=2\pi(678.1-i1.6)$~rad/s exhibits a remarkably small imaginary part and consequently a much larger quality factor, $Q\simeq212$. This mode lies close to the edge of the band gap, where the density of states and group velocity of the available propagating states are strongly modified. As the frequency approaches the band-gap edge, the coupling between the localized or quasi-localized resonance and the propagating continuum can become strongly reduced, resulting in a narrow resonance and a long lifetime. The large $Q$ observed here therefore illustrates that high-$Q$ resonances do not necessarily require a frequency deep inside a complete band gap. A mode located close to a spectral edge can also exhibit weak radiative coupling because of the reduced availability of efficient radiation channels. 

\begin{table}[ht]
\centering
\caption{Real and imaginary parts of the QNM frequencies and corresponding quality factors. Modes are characterized following the labels used in Fig.~\ref{fig:fig2}.}
\label{tab:QNM_Q}
\setlength{\tabcolsep}{10pt}
\begin{tabular}{c c c c}
\hline
\hline
Mode &
$\Omega_n/(2\pi)$ [Hz] &
$\gamma_n/(2\pi)$ [Hz] &
$Q_n$ \\
\hline
(a) & 130.6  & $43.5$ & 1.50 \\
(b) & 279.9  & $49.3$ & 2.84 \\
(c) & 678.1  & $1.6$  & 211.91 \\
(d) & 810.2  & $4.1$  & 98.80 \\
(e) & 2065.9 & $21.3$ & 48.49 \\
\hline
\hline
\end{tabular}


\end{table}

\begin{figure*}
    \centering
    \includegraphics[width=\linewidth]{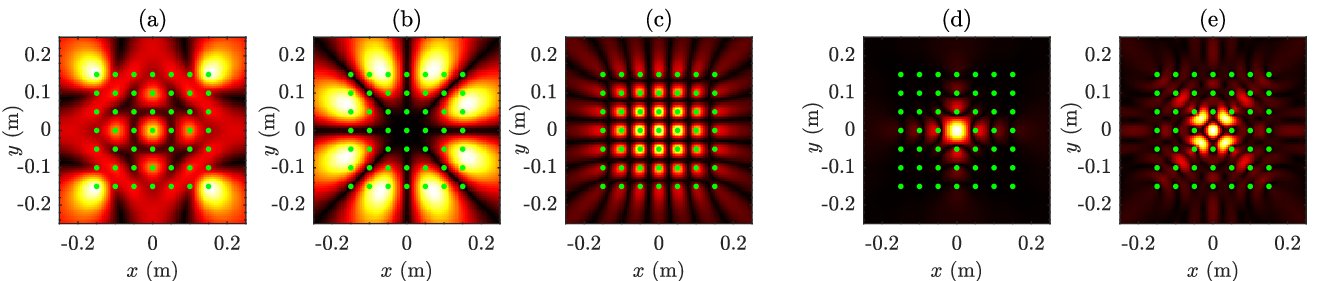}
    \caption{Comparison of the QNM displacement fields for a complete finite crystal (a, b, c) and for the corresponding structure containing a point defect (d, e). (a-e) Spatial distribution of the modal displacements associated with the QNM identified in Fig.~\ref{fig:fig1}(a) at $\omega_n=2\pi(130.6-i43.5)$ rad/s, $\omega_n=2\pi(279.9-i49.3)$ rad/s, $\omega_n=2\pi(678.1-i1.6)$ rad/s,  
    $\omega_n=2\pi(810.2-i4.1)$ rad/s, and $\omega_n=2\pi(2065.9-i21.3)$ rad/s respectively, showing the extended character of the mode throughout the finite crystal. The comparison illustrates the distinct character of the QNMs associated with propagating states in the pass bands and localized defect states within the band gaps.    }
    \label{fig:fig2}
\end{figure*}

Now the modal displacement of a QNM can be obtained from the corresponding null vector. Since the determinant is a complex-valued function of the complex frequency, the zero can be located numerically in the complex-frequency plane using, for example, Muller's method~\cite{atkinson89, Cortes26}. Starting from a set of initial complex-frequency estimates, the method constructs a quadratic interpolation of the function and iteratively determines the root of this interpolant, allowing the complex zero to be located without an explicit derivative of the determinant. This procedure is particularly convenient for the present non-self-adjoint problem, where the QNM frequencies are intrinsically complex.

When the QNM frequency $\omega_n$ has been determined with the required accuracy, the corresponding modal amplitudes are obtained by solving the homogeneous system
\begin{equation}
\sum_{\beta=1}^{N}\mathcal{M}_{\alpha\beta}(\omega_n)\psi_{\mathrm{e}}({\bf R}_{\beta})=0,
\quad 1 \leq \alpha \leq N
\end{equation}
where $\{\psi_{\mathrm{e}}({\bf R}_{1}),\ldots,\psi_{\mathrm{e}}({\bf R}_{N}) \}$ is the null vector of $\mathcal{\bm{M}}(\omega_n)$.  
The complete QNM displacement field is subsequently reconstructed by superposing the contributions radiated by all point masses through the outgoing Green's function. Because QNMs are solutions of a homogeneous problem, their overall amplitude is arbitrary; consequently, the resulting modal displacement is normalized. 
This procedure provides both the complex resonance frequency and its associated spatial mode profile, enabling the localization and spatial structure of the QNMs to be analyzed directly.

Based on both the QNM frequency and mode shape, the spectrum of a finite open periodic array contains modes with qualitatively different spatial characteristics. We therefore classify the QNMs according to their modal displacement, together with their position in the complex-frequency plane. This distinction is essential because the complex eigenfrequency alone does not uniquely determine the physical nature of a mode. In particular, the finite structure can support resonances associated with the bulk of the crystal, with its external boundary, or with the outgoing-wave character of the open-system eigenvalue problem. Based on the spatial distribution of the reconstructed modal displacement, we identify two main classes of QNMs: bulk-like modes and boundary-localized modes.

\paragraph{Bulk-like modes.} Bulk-like QNMs are characterized by a displacement field that extends throughout the interior of the finite structure and resembles the spatial structure of the propagating states of the corresponding infinite periodic system. At low long wavelengths, the bulk-like QNMs recover the behavior of the corresponding homogenized medium, since the wavelength is sufficiently large for the microscopic details of the point distribution to be effectively averaged out. Their real frequencies predominantly occur in the pass bands of the bulk band structure. These modes couple to the exterior through the finite boundaries and therefore acquire a finite negative imaginary part, which quantifies their radiative decay. Their spatially extended character provides a direct connection between the QNM spectrum of the finite structure and the bulk bands obtained from the self-adjoint periodic problem. As an example of Bulk-like mode we show in Figs.~\ref{fig:fig2}(a,c) the complex frequency map of the minimum eigenvalue of the Foldy-Lax matrix and the modal displacement for $\omega_n=2\pi(130-i43.5)$ rad/s and $\omega_n=2\pi(678.1-i1.6)$ rad/s respectively.

\paragraph{Boundary-localized modes.} A second class consists of modes whose displacement is preferentially concentrated close to the outer boundary of the finite structure, with comparatively weak amplitude in its interior. These modes exhibit a spatial character reminiscent of whispering-gallery resonances~\cite{Baryshnikov04, Nazmitdinov01} and are primarily associated with the finite geometry and its boundary rather than with the bulk spectrum. 
Their presence is therefore a finite-size effect and should be distinguished from genuine defect states localized around an internal structural perturbation. As an example of boundary-localized mode we show in Fig.~\ref{fig:fig2}(b) the complex frequency map of the minimum eigenvalue of the Foldy-Lax matrix and the modal displacement for $\omega_n=2\pi(279.9-i49.3)$ rad/s.

Having established the QNM spectrum of the periodic system, we next investigate how a localized structural perturbation modifies this spectrum. In particular, the introduction of a point defect breaks the translational symmetry with respect to a perfectly periodic configuration and can generate localized states at frequencies for which the full crystal supports no propagating bulk modes. Within the QNM framework, as shown in Fig.~\ref{fig:fig1}(b), these defect states appear as isolated complex-frequency resonances which radiative linewidth can be directly characterized by the imaginary parts of the QNM frequencies. Now we investigate its spatial localization character from the corresponding eigenvectors.

\paragraph{Defect-induced modes.}
When a point defect is introduced into the otherwise periodic system, an additional class of QNMs can emerge within the bulk band gaps. These defect-induced modes originate from the local breaking of translational symmetry and are characterized by a displacement field strongly localized around the defect. In contrast to bulk-like modes, their frequencies do not belong to the propagating bands of the full crystal, while, unlike boundary-localized modes, their localization is determined by the internal defect rather than by the external boundary. The real part of the QNM frequency therefore lies within a bulk band gap, where the surrounding crystal supports only evanescent fields. Consequently, the modal displacement decreases rapidly away from the defect, reflecting the exponential spatial confinement associated with the band gap. As two examples of defect-induced modes, we show in Figs.~\ref{fig:fig2}(d,e) the modal displacements for $\omega_n=2\pi(810.2-i4.1)$ rad/s and $\omega_n=2\pi(2065.9-i21.3)$ rad/s respectively. The defect-induced QNM can therefore be distinguished from the other classes by the simultaneous occurrence of a frequency inside a bulk band gap, strong localization around the defect, and a small imaginary part that decreases with increasing system size.

The first defect mode, at $\omega_n=2\pi(810.2-i4.1)$~rad/s, lies within the complete band gap of the crystal and exhibits a relatively small imaginary part, corresponding to a high quality factor of $Q\simeq98.8$. Its frequency is spectrally isolated from the propagating bulk modes, and the surrounding array therefore strongly suppresses the coupling of the defect state to propagating channels. The finite imaginary part nevertheless indicates that radiation is not completely suppressed, since the evanescent field associated with the defect mode can extend towards the external boundaries and provide a residual leakage channel.

The second defect mode, at $\omega_n=2\pi(2065.9-i21.3)$~rad/s, with $Q\simeq48.5$, lies outside the complete band gap and within a propagating band of the crystal. In contrast to the first defect state, this mode can directly couple to extended bulk states of the crystal, providing additional pathways for energy to escape from the localized region. The larger magnitude of $\gamma_n$ is therefore consistent with enhanced radiative coupling and a shorter modal lifetime. The comparison between the two defect modes illustrates the role of the spectral environment in determining the $Q$ factor: spatial localization alone does not determine the radiative lifetime, which is also governed by the availability and strength of the channels to which the localized state can couple.


An additional characteristic of the defect-induced QNM is the dependence of its imaginary part on the size of the finite system. As the dimensions of the array are increased, the defect is progressively separated from the external boundary, and the evanescent tail of the localized mode reaching the radiative region becomes increasingly small. Consequently, the coupling of the defect state to the exterior is reduced and the magnitude of $\gamma_n$ decreases. In the limit of an infinitely extended crystal, the defect state becomes an exponentially localized trapped state within the band gap, for which $\gamma_n\rightarrow0$ and the QNM approaches a real eigenfrequency. For a finite system, the mode remains a QNM because of its residual coupling to the radiative continuum, but its quality factor increases with the distance between the defect and the boundary. This finite-size dependence provides an additional confirmation that the complex frequency associated with the defect originates from radiative leakage rather than from intrinsic material losses.

For a defect state localized within a band gap, the field outside the defect region is evanescent, so that its amplitude at the boundary decreases approximately as $\exp(-\kappa L/2)$, where $\kappa$ is the attenuation constant in the band gap and $L/2$ characterizes the distance between the defect and the boundary. Since the radiative decay rate is controlled by the squared amplitude reaching the boundary, one expects an approximately exponential dependence of the QNM linewidth, $|\gamma_n| \propto e^{-\kappa L}$, and consequently an exponential increase of the quality factor with supercell size. This behavior provides a direct connection between the evanescent localization of the defect mode in the bulk band gap and the radiative lifetime of its finite-system QNM ~\cite{Romero10_1, Romero10_2, Romero10, Romero11}.

This classification and analysis highlights the importance of combining the spectral information contained in the complex QNM frequencies with the spatial information provided by their eigenvectors. In particular, bulk-like, boundary-localized, and defect QNMs can occupy overlapping regions of the complex-frequency plane and therefore cannot, in general, be distinguished solely from their eigenfrequencies. The modal displacement provides the necessary additional information to identify the physical origin and spatial character of each resonance, and the complex-frequency provides information about the lifetime and the resonance frequency.


\subsection{Stealthy hyperuniform systems}

To investigate the influence of aperiodic spatial correlations on the QNM spectrum, we consider three two-dimensional stealthy hyperuniform point configurations containing $N=49$ points distributed within a square region of side length $L=7a$. The configurations are generated using a dedicated numerical algorithm that imposes the stealthy constraint on the structure factor, $S(\mathbf{q})=0$, for wavevectors satisfying $0<|\mathbf{q}|\leq K$ \cite{Romero19, Kuznetsova24}. The degree of stealthiness is quantified by the dimensionless parameter $\chi$, which measures the fraction of configurational degrees of freedom constrained by this condition. Three configurations with increasing values $\chi=0.19$, $0.46$, and $0.56$ are considered, thereby allowing us to systematically investigate the effect of increasing spatial correlations on the spectral and resonant properties of the resulting structures. Moreover, the three configurations have identical numbers of points and occupy the same area as in the periodic case facilitating the comparison. The differences between the stealthy hyperuniform point patterns arise solely from the increasing degree of stealthy constraints.

\begin{figure*}
    \includegraphics[width=\textwidth]{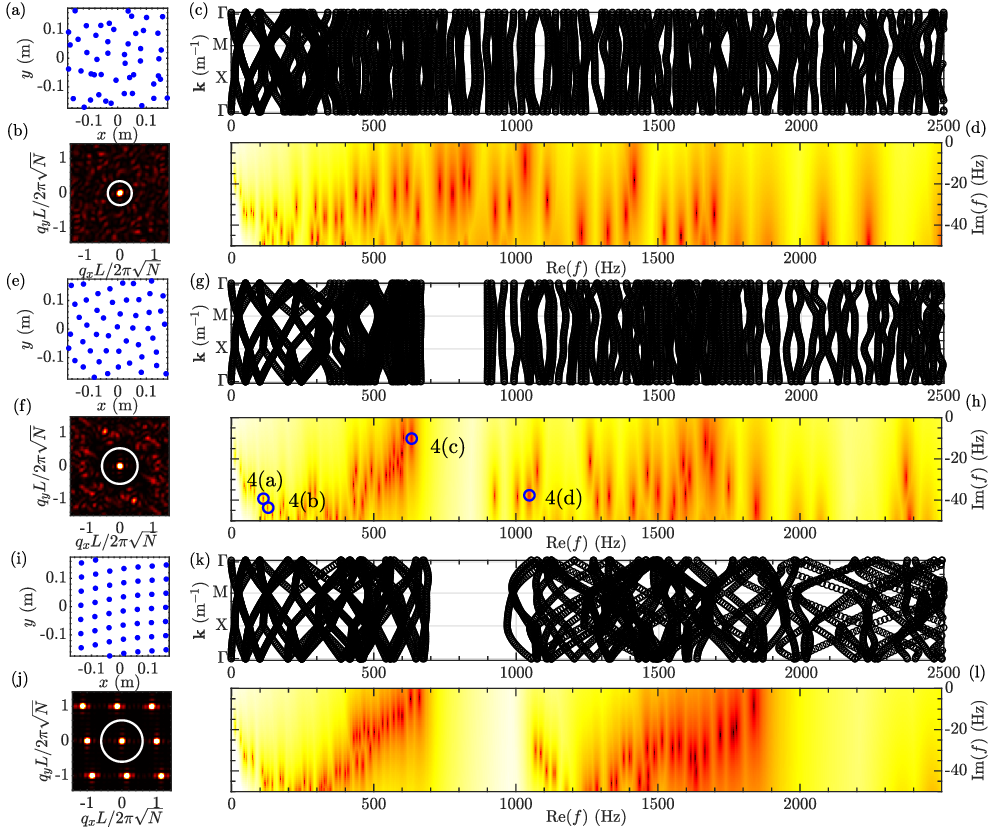}
    \caption{ Characterization of three stealthy hyperuniform configurations with increasing stealthiness parameter $\chi$. For $\chi=0.19$, $0.46$, and $0.56$, the corresponding results are shown in the first $2\times2$ block of figures (a-d), second block (e-h), and third block (k-l), respectively. Figures (a), (e), and (i) show the point distributions. Panels (b), (f), and (j) present the corresponding structure factors $S(\mathbf{q})$, where the white solid circle indicates the stealthy region in reciprocal space, $|\mathbf{q}|\leq K$, within which the structure factor is suppressed, highlighting the progressive extension of the stealthy region with increasing $\chi\simeq\frac{(KL)^2}{8\pi N}$. Axis are normalized by $2\pi/(L/\sqrt{N})$. Figures (c), (g), and (k) show the band structures calculated using the PWE method for the corresponding supercells. Finally, Figures (d), (h), and (l) display the minimum eigenvalue of the Foldy-Lax matrix in the complex-frequency plane, represented on a logarithmic scale, providing the QNM spectral characterization of each finite hyperuniform realization. The comparison illustrates the evolution of the structural correlations and their manifestation in both the real-frequency band structure and the complex-frequency spectrum as the degree of stealthiness is increased.
}
\label{fig:fig3}
\end{figure*}

The three configurations considered here illustrate the progressive modification of the spatial correlations as the stealthiness parameter is increased from $\chi=0.19$ to $\chi=0.46$ and $\chi=0.56$, as shown in Figs.~\ref{fig:fig3}(a, e, i), respectively. As expected, increasing $\chi$ enlarges the region of reciprocal space over which the structure factor is suppressed, as evidenced in Figs.~\ref{fig:fig3}(b, f, j) by the increasing radius $K$ of the stealthy region in the structure factor, indicated by the white circle. Although none of the configurations exhibits translational periodicity, the point distributions become progressively more constrained as $\chi$ increases, demonstrating that the evolution from weak to strong stealthiness reflects a progressive increase in the spatial correlations of the aperiodic structure. The corresponding PWE calculations in Figs.~\ref{fig:fig3}(c, g, k) reveal how these correlations are manifested in the spectral properties of the periodically repeated supercell. In particular, the band structures evolve systematically with $\chi$, providing a conventional real-frequency description of the collective wave propagation supported by each hyperuniform configuration.

\begin{figure*}
    \includegraphics[width=\textwidth]{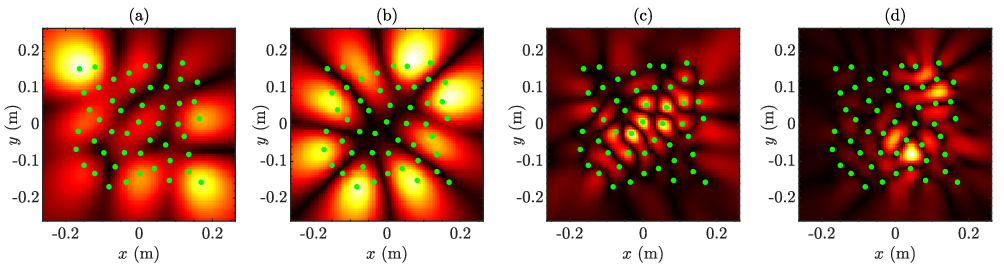}
\caption{%
Modal displacement fields of representative QNMs of the finite stealthy hyperuniform structure with $\chi\simeq0.46$. 
The complex eigenfrequencies are (a) $\omega_n=2\pi(111.8-i39.1)$ rad/s, (b) $\omega_n=2\pi(128.7-i43.7)$ rad/s, (c) $\omega_n=2\pi(633.4-i10.2)$ rad/s, and (d) $\omega_n=2\pi(1047.7-i37.6)$ rad/s as indicated in Fig.~\ref{fig:fig3}(h). 
(a) and (c) show bulk-like QNMs, characterized by spatially extended modal displacement fields in the structure. (b) corresponds to a boundary-localized QNM, with the modal amplitude concentrated near the external boundary of the finite structure. (d) shows an intrinsically localized QNM, whose displacement is confined to a localized region of the hyperuniform configuration in the absence of an externally introduced defect.
}
    \label{fig:fig4}
\end{figure*}
A complementary picture is obtained from the complex-frequency spectra in Figs.~\ref{fig:fig3}(d, h, l). 
The evolution of this complex spectrum with $\chi$ demonstrates that the structural correlations imposed by stealthy hyperuniformity are directly reflected in the resonant response of the finite structure. In particular, the QNM frequency distribution provides information about mode lifetimes and spatially localized resonances that cannot be inferred from the structure factor alone. Thus, the comparison between the three descriptions---$S(\mathbf{q})$, the PWE band structure, and the complex QNM spectrum---establishes a direct connection between the reciprocal-space constraints defining stealthy hyperuniformity and the resonant properties of its finite realizations, as we explain in the following.

As $\chi$ increases, the enhancement of the spatial correlations is accompanied by a pronounced reorganization of the spectral properties of the stealthy hyperuniform structures. In the PWE calculations, a frequency interval with a strongly reduced density of states progressively develops and becomes increasingly well defined for the configurations for $\chi\gtrsim0.4$ as shown in the examples with $\chi=0.46$ and $\chi=0.56$. Since the structures are aperiodic and do not possess a conventional Brillouin zone, this feature should not be interpreted as a Bragg band gap in the usual crystalline sense. We therefore refer to it as a \emph{hyperuniform spectral gap} or pseudogap. Its emergence demonstrates that the suppression of long-wavelength density fluctuations associated with increasing $\chi$ can induce a pronounced spectral separation even in the absence of translational periodicity. The effect is particularly remarkable because it occurs as a consequence of the collective spatial correlations of the point pattern rather than from conventional periodic Bragg scattering.

The QNM spectra provide a complementary and particularly revealing view of this evolution. For the weakly stealthy configuration, $\chi=0.19$, the complex-frequency resonances are distributed over a broad range of real frequencies, without a pronounced depletion corresponding to the spectral structure observed at larger $\chi$. As the stealthiness increases, the QNMs progressively organize into frequency regions associated with the propagating bands originated from the periodicity of the supercell made by the stealthy hyperuniform structure. In particular, the QNM spectrum is strongly depleted in the frequency interval corresponding to the hyperuniform spectral gap, while a significant number of resonances are found within the pass bands. This correspondence between the real-frequency spectral organization obtained from PWE and the distribution of complex-frequency QNMs demonstrates that the collective spatial correlations imposed by stealthy hyperuniformity are directly inherited by the resonant spectrum of the finite open system. Importantly, the QNM analysis additionally reveals the radiative character of these states through their imaginary frequencies, providing information about resonance lifetimes that is not accessible from the PWE band structure alone. This correspondence is nontrivial because the PWE calculation describes the periodically repeated supercell, whereas the QNM calculation probes a single finite and open realization; nevertheless, the spectral gap emerging from the collective hyperuniform correlations is clearly reflected in the complex-frequency distribution of the finite structure.

For the stealthy hyperuniform configuration with ${\chi\simeq 0.46}$, the point pattern occupies an intermediate structural regime between a periodic system and a conventional disordered medium. Although translational periodicity is absent, the strong spatial constraints imposed by stealthy hyperuniformity result in pronounced structural correlations and suppress long-wavelength density fluctuations. The resulting aperiodic structure is therefore neither periodic nor uncorrelated, providing a distinct regime in which to investigate how nontrivial spatial correlations are manifested in the QNM spectrum. Generally speaking, three qualitatively different families of QNMs can be identified from their modal displacement fields: bulk-like, boundary-localized, and intrinsically-localized QNMs.

\begin{figure*}
    \centering
    \includegraphics[width=\linewidth]{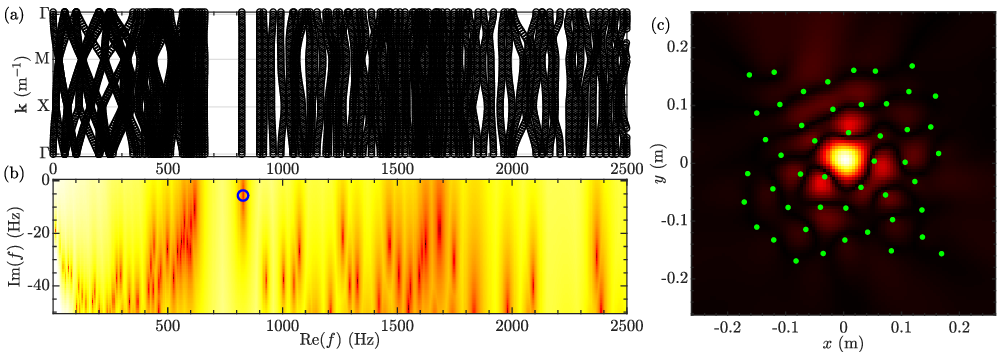}
    \caption{%
Defect-induced localized state in a stealthy hyperuniform structure with $\chi\simeq0.46$. 
(a) PWE band structure of the periodically repeated hyperuniform supercell with a point defect introduced by removing the central point mass. The defect gives rise to an approximately flat band within the hyperuniform pseudogap. 
(b) Complex-frequency QNM spectrum of the corresponding finite open realization. The isolated QNM within the pseudogap is indicated, with its position in the complex-frequency plane characterizing both the resonance frequency and its radiative linewidth. Blue open circle corresponds to the QNM for this particular localized mode, $\omega_n=2\pi(827.8-5.6i)$ rad/s.
(c) Corresponding modal displacement field, showing the strong spatial localization of the QNM around the position of the removed point mass. The localized character of the mode demonstrates the formation of a defect resonance within the pseudogap of the hyperuniform background.
}
    \label{fig:fig5}
\end{figure*}


\paragraph{Bulk-like modes.} In the low frequency regime, as shown in Fig.~\ref{fig:fig4}(a), the QNMs exhibit the behavior expected from the effective, homogenized medium. The corresponding modes are extended over the structure and can be classified as bulk modes. In this long-wavelength regime, \st{the} waves do not strongly resolve the microscopic aperiodic arrangement of the point masses, and the modal response is therefore governed predominantly by the effective properties of the hyperuniform medium. As the frequency increases, the QNMs progressively become sensitive to the underlying correlated point pattern. A second family of collective bulk modes emerges, Fig.~\ref{fig:fig4}(c), characterized by extended displacement fields that nevertheless exhibit pronounced spatial fluctuations reflecting the specific hyperuniform realization. These modes are neither conventional Bloch modes nor modes of an uncorrelated random medium; rather, they arise from the collective multiple-scattering response of the correlated aperiodic structure.

\paragraph{Boundary-localized modes.} Figure \ref{fig:fig4}(b) shows a state that arises from the truncation of the aperiodic medium at its external boundary and is characterized by a modal displacement field\st{s} concentrated preferentially near the edges of the finite sample. The localization of this type of mode is not associated with a particular local configuration of the point pattern, but rather with the termination of the hyperuniform medium. Its presence therefore highlights the importance of the finite geometry and open boundaries in the QNM spectrum, and further distinguishes the finite system from its periodically repeated supercell description.

\paragraph{Intrinsically-localized modes.} At higher frequencies within the first propagating band, intrinsically localized modes emerge. These states are spatially confined to specific regions of the hyperuniform configuration despite the absence of an externally introduced defect as shown in Fig.~\ref{fig:fig4}(d). Their localization therefore results from the intrinsic spatial organization of the aperiodic point pattern and the associated multiple-scattering interference. This behavior is particularly significant because it demonstrates that localization can emerge in a stealthy hyperuniform structure without relying on conventional periodic band-gap confinement or on a deliberately introduced point defect. The QNM description thus reveals a hierarchy of modal responses, from homogenized behavior at long wavelengths, through collective modes sensitive to the correlated aperiodic structure, to intrinsically localized resonances.

The spatial distributions of the modes at ${\omega_n=2\pi(633.4-i10.2)}$~rad/s [Fig.~\ref{fig:fig4}(c)] and ${\omega_n=2\pi(1047.7-i37.6)}$~rad/s  [Fig.~\ref{fig:fig4}(d)] exhibit markedly different characteristics. For the mode at $633.4$~Hz, the maximum displacement is predominantly concentrated around the positions occupied by the point masses. This behavior indicates that the mode is strongly coupled to the local scattering elements and can therefore be regarded as a particle-centered collective mode. Despite its localization around the scatterers, the mode extends over a substantial portion of the structure, reflecting its collective character rather than that of an isolated localized resonance. 

In contrast, the QNM at $1047.7$~Hz exhibits a qualitatively different spatial character, with its largest displacement occurring predominantly in the regions between neighboring point masses rather than at the positions of the scatterers. This interstitial character can be understood as a consequence of the collective interference of the waves scattered by the surrounding masses. In an aperiodic hyperuniform structure, the correlated spatial arrangement of the scatterers can organize the field into collective patterns with distinct nodal structures. The contrast between the particle-centered and interstitial modes illustrates that the QNM spectrum contains information not only about the spectral distribution of resonances, but also about how the wave field is organized relative to the underlying hyperuniform point pattern.

The presence of intrinsic localized states in a stealthy hyperuniform structure does not preclude the controlled creation of additional localized states through an intentional defect. To investigate this possibility, we introduce a point defect in the central region of the $\chi\simeq0.46$ hyperuniform configuration by removing one of the point masses. 

\paragraph{Defect-induced modes.} The corresponding supercell calculation reveals the emergence of an approximately flat spectral branch within the hyperuniform pseudogap [see Fig.~\ref{fig:fig5}(a)]. The weak dispersion of this branch indicates that the associated state is spatially confined and has only weak coupling to the extended collective modes of the surrounding hyperuniform structure. The QNM calculation for the corresponding finite realization provides a complementary characterization: an isolated complex-frequency resonance is found within the pseudogap [see Fig.~\ref{fig:fig5}(b)], and its modal displacement is strongly concentrated around the position of the removed mass [see Fig.~\ref{fig:fig5}(c)]. The defect therefore acts as a localized resonant state embedded in a frequency interval where the extended states of the hyperuniform background are strongly suppressed.

The QNM spectrum of the $\chi=0.46$ stealthy hyperuniform configuration further illustrates the strong influence of the hyperuniform spectral environment on the radiative properties of the modes (see Tab.~\ref{tab:QNM_chi046}). The two low-frequency modes, at $111.8$ and $128.7$~Hz, exhibit relatively large imaginary parts, corresponding to $Q\simeq1.43$ and $Q\simeq1.47$, respectively. These broad resonances are consistent with strongly radiative bulk- and boundary-like states, for which efficient coupling to the external continuum results in short modal lifetimes. The mode at $633.4$~Hz presents a substantially smaller linewidth, with $Q\simeq31$, indicating a significant reduction in its coupling to the radiative channels. The mode at $1047.7$~Hz, in contrast, has a broader linewidth, with $Q\simeq14$, illustrating that the radiative lifetime is not determined solely by frequency but depends strongly on the spatial character of the mode and its overlap with the available radiation channels.

\begin{table}[ht]
\centering
\caption{Real and imaginary parts of the QNM frequencies and corresponding quality factors for the SHU configuration with $\chi=0.46$. Labels of the modes follows the one used in Fig.~\ref{fig:fig4}.}
\label{tab:QNM_chi046}
\setlength{\tabcolsep}{10pt}
\begin{tabular}{cccc}
\hline
\hline
Mode &
$\Omega_n/(2\pi)$ [Hz] &
$\gamma_n/(2\pi)$ [Hz] &
$Q_n$ \\
\hline
(a) & 111.8  & $39.1$ & 1.43 \\
(b) & 128.7  & $43.7$ & 1.47 \\
(c) & 633.4  & $10.2$ & 31.05 \\
(d) & 1047.7 & $37.6$ & 13.93 \\
defect & 827.8  & $5.6$  & 73.91 \\
\hline
\hline
\end{tabular}

\end{table}

The defect-induced state in the hyperuniform structure differs qualitatively from the conventional defect mode of a periodic system. In a periodic system, a point defect introduced inside a well-defined band gap produces a localized state whose frequency is isolated from the propagating Bloch bands of the surrounding crystal. The translational symmetry of the background strongly constrains the available coupling channels, and, for a sufficiently large finite crystal, the defect mode can become exponentially decoupled from the external boundary. As a consequence, its QNM frequency can approach the real axis as the size of the crystal is increased, resulting in a high-$Q$ resonance \footnote{The small imaginary part of the QNM frequency should not be interpreted as a direct measure of spatial localization. Rather, it quantifies the temporal decay rate of the resonance and, consequently, its coupling to the available radiation channels. Spatial localization can nevertheless lead to a small $\gamma_n$ when it suppresses the overlap of the mode with these channels. For a defect state lying inside a complete band gap, for example, the absence of propagating bulk states forces the field to decay spatially away from the defect, reducing its amplitude at the external boundaries and thereby suppressing radiative leakage.}. In contrast, the pseudogap of a hyperuniform structure is not a conventional band gap associated with the absence of Bloch states. It originates from the collective correlations of an aperiodic configuration and is accompanied by a much richer spectrum of collective and localized states. The defect state is therefore embedded in a structurally heterogeneous environment rather than in a perfectly periodic background.

This distinction has important consequences for the radiative properties of the defect resonance. Although the defect-induced mode lies inside the hyperuniform pseudogap and is strongly localized around the position of the removed mass, the aperiodic background can provide additional channels for coupling between the defect state and the radiative continuum. Local structural fluctuations can mix the defect mode with nearby collective or intrinsically localized QNMs, while the absence of translational symmetry removes the constrains coupling in the periodic system.  Consequently, as shown in Tab.~\ref{tab:QNM_chi046} the defect QNM at $\omega_n=2\pi(827.8-i5.6)$~rad/s in the hyperuniform structure exhibit larger $\left|\gamma_n\right|$ than the corresponding defect mode in a periodic system, despite both modes being localized and lying within a spectral gap or pseudogap. The imaginary part of the QNM frequency therefore provides a direct measure of the different radiative environments experienced by the two defect states. In this sense, the hyperuniform pseudogap suppresses the density of extended states but does not necessarily suppress all radiative coupling channels, highlighting the distinction between spectral localization and radiative isolation. The finite value of $\gamma_n$ therefore indicates residual coupling between the defect state and the external continuum, mediated by the finite size and the aperiodic scattering environment. The resulting resonance can consequently be regarded as a weakly radiative localized state whose lifetime is controlled by the interplay between defect localization, the hyperuniform pseudogap, and the available radiation channels.


\section{Discussion}

The results presented above demonstrate that the QNM spectrum provides a complementary characterization of hyperuniform structures that goes beyond the conventional description based on the structure factor and the band structure of a periodically repeated supercell. The structure factor quantifies the suppression of long-wavelength density fluctuations and therefore provides a direct measure of the stealthy constraints imposed on the point distribution. The PWE calculation, in turn, reveals how these structural correlations are manifested in the real-frequency spectrum of the corresponding periodic supercell. However, neither quantity directly describes the resonant response of a finite and open realization. The QNM formulation addresses this limitation by providing a complex-frequency spectrum in which both the resonance frequency and the radiative decay rate are simultaneously encoded.

For the stealthy hyperuniform configurations investigated here, increasing the stealthiness parameter $\chi$ produces a progressive enhancement of the spatial correlations while preserving the absence of translational periodicity. This evolution is accompanied by a pronounced reorganization of the spectral response. At low $\chi$, the QNMs of the finite structure are broadly distributed over the real-frequency axis, reflecting the absence of a well-defined spectral separation between propagating and non-propagating frequency ranges. As $\chi$ increases, a spectral pseudogap progressively develops in the PWE calculation. Although this feature resembles a band gap of a periodic system, it should not be interpreted as a conventional Bragg gap because the underlying structure is aperiodic and does not possess a Brillouin-zone description in the usual sense. Instead, the pseudogap results from the collective spatial correlations imposed by the stealthy-hyperuniform constraints.

An important observation is that the QNM spectrum of the finite realization reflects this spectral organization despite the fundamentally different nature of the other two calculations. On the one hand, the PWE method considers the periodic continuation of a selected hyperuniform configuration, yielding a self-adjoint problem with real eigenfrequencies. On the other hand, the QNM calculation considers a single finite and open realization and yields a non-self-adjoint problem with complex eigenfrequencies. Nevertheless, the QNMs are predominantly concentrated within the frequency intervals identified as propagating bands by the PWE calculation, while a strong depletion of QNMs is observed in the hyperuniform pseudogap. This correspondence demonstrates that the collective correlations responsible for the spectral pseudogap survive the transition from the periodically repeated supercell to the finite open structure. At the same time, the complex-frequency description provides additional information through $\operatorname{Im}(\omega)$, which quantifies the radiative linewidth and therefore the lifetime of each resonance.

The spatial structure of the QNMs further reveals a hierarchy of modal regimes within the first propagating band. In the low frequency regime, the modes are well described by the effective-medium response and consist primarily of extended bulk modes and modes associated with the external boundary. In this long-wavelength regime, the wavelength is sufficiently large that the individual point scatterers are only weakly resolved. At intermediate frequencies, collective bulk modes emerge. These states remain spatially extended but increasingly reflect the specific correlated aperiodic arrangement of the hyperuniform configuration, indicating that the wave begins to resolve the microscopic structure while maintaining collective character. At higher frequencies within the propagating band, intrinsically localized modes appear. These states are localized around specific regions of the hyperuniform configuration without the introduction of an externally imposed defect. Their existence highlights an important distinction between hyperuniform and periodic systems: localization can arise from the intrinsic spatial organization of an aperiodic hyperuniform structure, rather than solely from a band gap or an intentionally introduced defect.

The introduction of a controlled point defect provides a further distinction between intrinsic localization in hyperuniform structures and conventional defect states in periodic systems. Removing a point mass from the central region of a hyperuniform configuration produces an approximately flat branch inside the spectral pseudogap of the supercell calculation. The corresponding QNM is strongly localized around the defect position, demonstrating that a localized resonance can be deliberately introduced within the frequency range in which the extended states of the hyperuniform background are strongly suppressed. This behavior is reminiscent of the defect modes of periodic systems, but the underlying radiation physics is different. In a periodic system, a defect embedded in a well-defined band gap is surrounded by a translationally ordered medium, and the coupling to propagating Bloch channels is strongly constrained. In a hyperuniform structure, by contrast, the pseudogap is generated by correlated aperiodic order rather than by translational symmetry. Consequently, the defect mode can remain coupled to collective and localized states of the hyperuniform background and, through them, to the external radiative continuum.

This difference is particularly evident in the imaginary part of the defect QNM frequency. The localization of the defect mode does not by itself guarantee a vanishing radiative linewidth. In the finite hyperuniform structure, the evanescent field surrounding the defect can couple to the external boundary, while the aperiodic background can provide additional mode-mixing pathways that are absent or strongly constrained in a periodic system. The resulting radiative leakage is therefore determined not only by the localization length of the defect state, but also by the availability of scattering channels in the surrounding hyperuniform medium. The QNM framework makes this distinction explicit: two defect states with comparable spatial localization and similar real frequencies can nevertheless possess substantially different quality factors because of their different coupling to the radiative continuum. This illustrates the advantage of analyzing the complex spectrum rather than relying exclusively on the real-frequency band structure.

The finite size of the structure provides an additional control parameter for the defect-state linewidth. For a localized state whose frequency lies within the hyperuniform pseudogap, the coupling to the external boundary is expected to decrease as the distance between the defect and the boundary increases. In the regime where the defect mode is exponentially localized, this contribution to the radiative leakage is expected to decrease approximately exponentially with the system size. Consequently, the imaginary part of the defect QNM frequency should progressively approach zero as the finite realization is enlarged, provided that the defect frequency remains within the pseudogap and that no additional radiative channels become available. This finite-size dependence distinguishes a genuinely localized defect resonance from an extended collective mode and provides an additional criterion for quantifying its radiative confinement.

\section{Concluding remarks}

In conclusion, we have investigated the QNM spectrum of finite mass-loaded plates with periodic and stealthy-hyperuniform point configurations. By combining a multiple-scattering formulation with a complex-frequency QNM analysis, we have characterized not only the resonance frequencies of the structures but also their spatial modal profiles. This approach provides a description of finite open realizations that is complementary to the conventional characterization of hyperuniform structures through the structure factor and to the band structures obtained from periodically repeated supercells.

For the stealthy-hyperuniform configurations, we have shown that increasing the stealthiness parameter $\chi$ progressively enhances the spatial correlations and produces a pronounced spectral pseudogap in the supercell PWE spectrum. Remarkably, this organization is also reflected in the QNM spectrum of a single finite realization: the resonances progressively organize within the propagating bands, while their density is strongly reduced in the pseudogap. Within the first propagating band, the QNM spatial profiles reveal three distinct regimes, evolving from homogenized bulk and edge modes at low frequencies, to collective bulk modes at intermediate frequencies, and finally to intrinsically localized states. These results demonstrate that the QNM spectrum provides a dynamical signature of the hidden correlations of hyperuniform structures that cannot be obtained from the structure factor alone.

The introduction of a point defect further demonstrates the ability of the QNM framework to distinguish controlled defect localization from intrinsic localization. Removing a point mass generates a localized state within the hyperuniform pseudogap, manifested as an approximately flat branch in the supercell spectrum and as an isolated complex-frequency QNM in the finite structure. Unlike a conventional defect state in a periodic system, however, the hyperuniform defect mode is embedded in an aperiodic correlated environment and can therefore couple to additional scattering and radiative channels. Its imaginary frequency consequently provides a direct measure of this coupling and of the resulting quality factor. The combination of spatial localization, spectral position, and radiative linewidth thus offers a more complete characterization of defect states than a real-frequency band structure alone. 

More broadly, these results establish the QNM spectrum as a complementary descriptor of hyperuniformity for finite open systems. While the structure factor quantifies the suppression of density fluctuations and the PWE spectrum identifies collective spectral features of the periodically repeated configuration, the complex-frequency spectrum additionally reveals how these structural correlations control resonance lifetimes, localization, and radiation. This perspective opens a route toward designing hyperuniform structures not only according to their reciprocal-space constraints, but also according to their desired resonant properties, including the spatial localization and quality factor of selected modes.

\section*{Acknowledgments}
MMS, LMGR, ML and VRG acknowledge grant PID2023-146237NB-I00 funded by MICIU/AEI/ 10.13039/501100011033. VFDP and VRG are funded by the European Union’s Horizon programme in the framework of the MSCA ``META-SONIC'' project under grant agreement No. 101202648.\\[.25cm]

\textit{Data Availability Statement} - The data that support the findings of this article are not publicly available. The data are available from the authors upon reasonable request.

\end{document}